\documentclass[aps,showkeys,showpacs,nofootinbib,floatfix]{revtex4}
\usepackage{amssymb}
\usepackage{amsmath}
\usepackage{graphicx}
\usepackage{multirow}
\def\({\left(}
\def\){\right)}
\def\[{\left[}
\def\]{\right]}

\begin{document}

\title{Constraints on $H_0$ from WMAP and baryon acoustic osillation measurements}
\author{Xue Zhang$^{1}$ \footnote{zhangxue@itp.ac.cn}
and Qing-Guo Huang$^{1,2,3}$ \footnote{huangqg@itp.ac.cn}}
\affiliation{
$^1$ CAS Key Laboratory of Theoretical Physics,\\
Institute of Theoretical Physics, \\
Chinese Academy of Sciences, Beijing 100190, China\\
$^2$ School of Physical Sciences, \\
University of Chinese Academy of Sciences,\\
No. 19A Yuquan Road, Beijing 100049, China\\
$^3$ Synergetic Innovation Center for Quantum Effects and Applications,
Hunan Normal University, Changsha 410081, China}
\date{\today}

\begin{abstract}
We report the constraints of $H_0$ obtained from Wilkinson Microwave Anisotropy Probe (WMAP)
combined with the latest baryonic acoustic oscillations (BAO) measurements.
We use the BAO measurements from the 6dF Galaxy Survey (6dFGS), the SDSS DR7 main galaxies sample (MGS),
the BOSS DR12 galaxies and the eBOSS DR14 quasars.
Adding the recent BAO measurements to the cosmic microwave background (CMB) data from WMAP,
we constrain cosmological parameters
$\Omega_m=0.298\pm0.005$, $H_0=68.36^{+0.53}_{-0.52}$ km s$^{-1}$ Mpc$^{-1}$,
$\sigma_8=0.8170^{+0.0159}_{-0.0175}$ in a spatially flat $\Lambda$ cold dark matter ($\Lambda$CDM) model,
and $\Omega_m=0.302\pm0.008$, $H_0=67.63\pm1.30$ km s$^{-1}$ Mpc$^{-1}$,
$\sigma_8=0.7988^{+0.0345}_{-0.0338}$ in a spatially flat $w$CDM model, respectively.
The combined constraint on $w$ from CMB and BAO in a spatially flat $w$CDM model is $w=-0.96\pm0.07$.
Our measured $H_0$ results prefer a value lower than 70 km s$^{-1}$ Mpc$^{-1}$,
consistent with the recent data on CMB constraints from Planck (2018),
but in $3.1\sim 3.5\sigma$ tension with local measurements of Riess et al. (2018)
in $\Lambda$CDM and $w$CDM framework, respectively.
Compared with the WMAP alone analysis, the WMPA+BAO analysis reduces the error bar by
75.4\% in $\Lambda$CDM model and 95.3\% in $w$CDM model.
\end{abstract}

\maketitle

\section{Introduction}

As we know, the Hubble constant $H_0$ are in tension
between the Cosmic Microwave Background (CMB) measurements from Planck \cite{Ade:2015xua,Aghanim:2018eyx}
and the type Ia supernova measurements from SH0ES \cite{Riess:2016jrr,Riess:2018}
(SNe, $H_0$, for the Equation of State of dark energy).
The value of $H_0$ can be directly obtained by the Hubble Space Telescope (HST)
and the CMB measurements (see \cite{Freedman:2010xv} for review on determining the Hubble constant).
Riess et al. (2016) \cite{Riess:2016jrr} reported $H_0=73.24 \pm 1.74$ km s$^{-1}$ Mpc$^{-1}$ (2.4\% precision) from Cepheids in the hosts of Type Ia supernovae (SNIa).
Recently Riess et al. (2018) \cite{Riess:2018} improves the precision to 2.3\%,
yielding $73.48 \pm 1.66$ km s$^{-1}$ Mpc$^{-1}$.
On the other hand, the Planck survey reported $H_0=67.27 \pm 0.66$ km s$^{-1}$ Mpc$^{-1}$
(0.98\%  precision; TT,TE,EE+lowP) in 2015 \cite{Ade:2015xua}
and $67.27 \pm 0.60$ km s$^{-1}$ Mpc$^{-1}$ (0.89\%  precision; TT,TE,EE+lowE) in 2018 \cite{Aghanim:2018eyx}.
There exist a $3.7\sigma$ tension between the new results of Planck and SH0ES.
Addison et al. (2016) \cite{Addison:2015wyg} have discussed the internal tension
inferred from the Planck data itself.
They have analyzed the Planck TT power spectra in detail and found that
the Hubble constant $H_0 = 69.7\pm1.7$ km s$^{-1}$ Mpc$^{-1}$ at the lower multipoles ($\ell < 1000$)
and $H_0 = 64.1\pm1.7$ km s$^{-1}$ Mpc$^{-1}$ at the higher multipoles ($\ell \geq 1000$).
The measured value of $H_0$ is much lower in the case of $\ell \geq 1000$.

At present it is difficult to explain the $H_0$ disagreement in the standard cosmological model.
The tensions among datasets could be due to some underestimated systematic error associated with the experiments.
Of course, we cannot exclude the possibility of new physics beyond the $\Lambda$CDM cosmology
\cite{Sola:2017znb,Zhao:2017cud,Qing-Guo:2016ykt,DiValentino:2016hlg,Pourtsidou:2016ico,Wang:2015wga,Wyman:2013lza},
so the additional crosschecks are expected.
In this paper we again call for another independent precise CMB measurements,
namely Wilkinson Microwave Anisotropy Probe (WMAP).
The 9-year WMAP reported a 3\% precision determination of $H_0=70.0 \pm 2.2$ km s$^{-1}$ Mpc$^{-1}$
in a spatially flat $\Lambda$CDM model \cite{Hinshaw:2012aka}.
On the other hand, Cheng et al. \cite{Cheng:2014kja} combined various BAO data sets
to get relatively tight constraints on $H_0 = 68.17^{+1.55}_{-1.56}$ km s$^{-1}$ Mpc$^{-1}$.
Combining the recent BAO measurements,
Wang et al. \cite{Wang:2017yfu} reported $H_0=69.13 \pm 2.34$ km s$^{-1}$ (3.38\% precision).
In addition, The Advanced LIGO and Virgo \cite{Abbott:2017xzu}
reported the a strong signal of GW170817 from the merger of a binary neutron-star system
and determined the Hubble constant $H_0=70.0^{+12.0}_{-8.0}$ km s$^{-1}$ Mpc$^{-1}$.
Addison et al. (2018) \cite{Addison:2017fdm} show that WMAP, Atacama Cosmology Telescope (ACT), South Pole Telescope (SPT) surveys, and primordial deuterium abundance constraints can be used together with some BAO data to provide the values of $H_0$,
which are $2.4\sim 3.1\sigma$ lower than SH0ES, independent of Planck.
Combining galaxy and Ly$\alpha$ BAO observations with the primordial deuterium abundance,
a value of $H_0=66.98 \pm 1.18$ km s$^{-1}$ Mpc$^{-1}$ has been estimated.
This value also have $3\sigma$ tension with local $H_0$ measurement.
These measurements, independent of SH0ES and Planck constraints,
seem to favor a lower $H_0$ value that is more consistent with Planck result.
See \cite{Busti:2014dua,Wang:2016iij,Yu:2017iju,Gomez-Valent:2018hwc,
Park:2018fxx,Park:2018tgj,Miao:2018zpw,Yang:2018euj,Yang:2018uae,Yang:2018qmz,Freedman:2017yms,Zhang:2018jfu}
for more literature on $H_0$.

Beyond the spatially flat standard cosmological model and without SH0ES and Planck measurements,
it would be interesting to study the Hubble constant constraints.
So, in this work we combine the BAO with WMAP measurements
to place constraints on $H_0$ in $\Lambda$CDM and $w$CDM cosmology.
The paper is organized as follows.
In section \ref{md}, we will introduce the model and data sets used in this work.
In section \ref{r}, we present our main results.
We conclude in section \ref{sd}.

\section{Model and Data}
\label{md}

In this paper we discuss the spatially flat $\Lambda$CDM and $w$CDM model,
first using the nine-year WMAP data only,
then combined with the additional BAO data sets.
We use the BAO measurements from the 6dFGS survey \cite{Beutler:2011hx},
the SDSS DR7 MGS \cite{Ross:2014qpa},
the BOSS DR12 (9-zbin) \cite{Wang:2016wjr},
and the eBOSS DR14 measurement \cite{Ata:2017dya}.
Their effective redshifts and constraints are listed in Table \ref{tab:BAO}.

\begin{table}[!htbp]
\centering
\begin{tabular}{c|c|c|c}
  \hline
  \hline
  Experiment & $z_\text{eff}$ & Measurement & Constraint \\
  \hline
  6dFGS & 0.106 & $r_d/D_V$ & $0.336 \pm 0.015$ Mpc \\
  \hline
  SDSS DR7 MGS & 0.15 & $D_V$ & $(664 \pm 25)(r_d/r_{d,\text{fid}})$ Mpc\\
  \hline
  BOSS DR12 (9zbin) & 0.31 & $D_A/r_d; H*r_d$ & $(6.29 \pm 0.14 )$ Mpc; $(11.55 \pm 0.70)\times 10^3 $ km/s\\
  &0.36 & $D_A/r_d$; $H*r_d$ & $ (7.09 \pm 0.16 )$ Mpc; $(11.81 \pm 0.50)\times 10^3 $ km/s\\
  &0.40 & $D_A/r_d$; $H*r_d$ & $ (7.70 \pm 0.16 )$ Mpc; $(12.12 \pm 0.30)\times 10^3 $ km/s\\
  &0.44 & $D_A/r_d$; $H*r_d$ & $ (8.20 \pm 0.13 )$ Mpc; $(12.53 \pm 0.27)\times 10^3 $ km/s\\
  &0.48 & $D_A/r_d$; $H*r_d$ & $ (8.64 \pm 0.11 )$ Mpc; $(12.97 \pm 0.30)\times 10^3 $ km/s\\
  &0.52 & $D_A/r_d$; $H*r_d$ & $ (8.90 \pm 0.12 )$ Mpc; $(13.94 \pm 0.39)\times 10^3 $ km/s\\
  &0.56 & $D_A/r_d$; $H*r_d$ & $ (9.16 \pm 0.14 )$ Mpc; $(13.79 \pm 0.34)\times 10^3 $ km/s\\
  &0.59 & $D_A/r_d$; $H*r_d$ & $ (9.45 \pm 0.17 )$ Mpc; $(14.55 \pm 0.47)\times 10^3 $ km/s\\
  &0.64 & $D_A/r_d$; $H*r_d$ & $ (9.62 \pm 0.22 )$ Mpc; $(14.60 \pm 0.44)\times 10^3 $ km/s\\
  \hline
  eBOSS DR14 & 1.52 & $D_V$ & $(3843 \pm 147)(r_d/r_{d,\text{fid}})$ Mpc\\
  \hline
  \hline
\end{tabular}
\caption{BAO distance measurements used in this work.}
\label{tab:BAO}
\end{table}

The angular diameter distance takes the form of
\begin{equation}
D_A (z) = \frac{1}{1+z} \int^z_0 \frac{d z'}{H(z')}.
\end{equation}
For the comoving sound horizon at the end of the baryon drag epoch $z_d$, we take $r_d \equiv r_s (z_d)$.
The volume-averaged effective distance $D_V$ is a combination of
the angular diameter distance $D_A(z)$ and Hubble parameter $H(z)$,
\begin{equation}
D_V(z)=\[(1+z)^2 D_A^2(z)\frac{cz}{H(z)}\]^{1/3}.
\end{equation}

We use the CosmoMC package \cite{Lewis:2002ah} to sample the parameter space.
Our analysis employs the same Markov Chain Monte Carlo (MCMC) formalism used in previous analyses
\cite{Abbott:2018wog,Brout:2018jch,Ade:2018gkx,Chen:2018dbv,Chen:2017ayg,Huang:2015vpa}.
We explore the WMAP-only and WMAP+BAO likelihood with MCMC simulations of the posterior distribution
for the six base parameters as given in Planck Collaboration \cite{Ade:2015xua,Aghanim:2018eyx}.
This approach naturally generates the likelihoods of parameters,
which are marginalized over all other fitting parameters.
The six basic parameters are
the baryon density today, $\Omega_b h^2$,
the cold dark matter density today, $\Omega_c h^2$,
100 $\times$ approximation to $r_*/D_A$, $100 \theta_{\rm MC}$,
the reionization optical depth, $\tau$,
the log power of the primordial curvature perturbations, 
$\ln(10^{10} A_s)$,
and the scalar spectrum power-law index, $n_s$.

\section{Result}
\label{r}

\begin{figure}[!htbp]
\centering
\includegraphics[width=.75\textwidth]{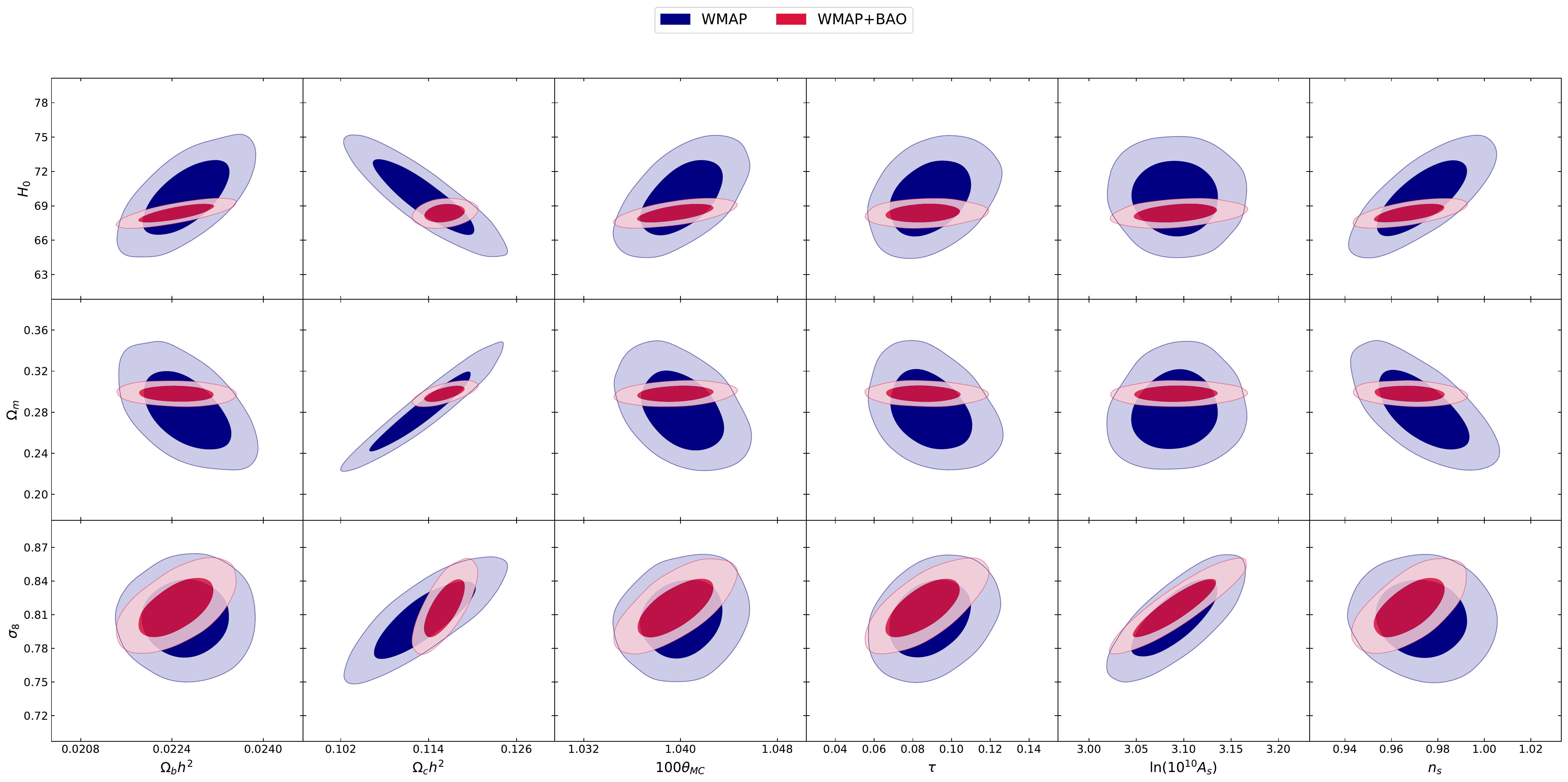}
\caption{Likelihood contours (68\% and 95\%) of cosmological parameters in a flat $\Lambda$CDM model derived from WMAP and WMAP + BAO respectively.}
\label{fig:lcdm}
\end{figure}

\begin{figure}[!htbp]
\centering
\includegraphics[width=.75\textwidth]{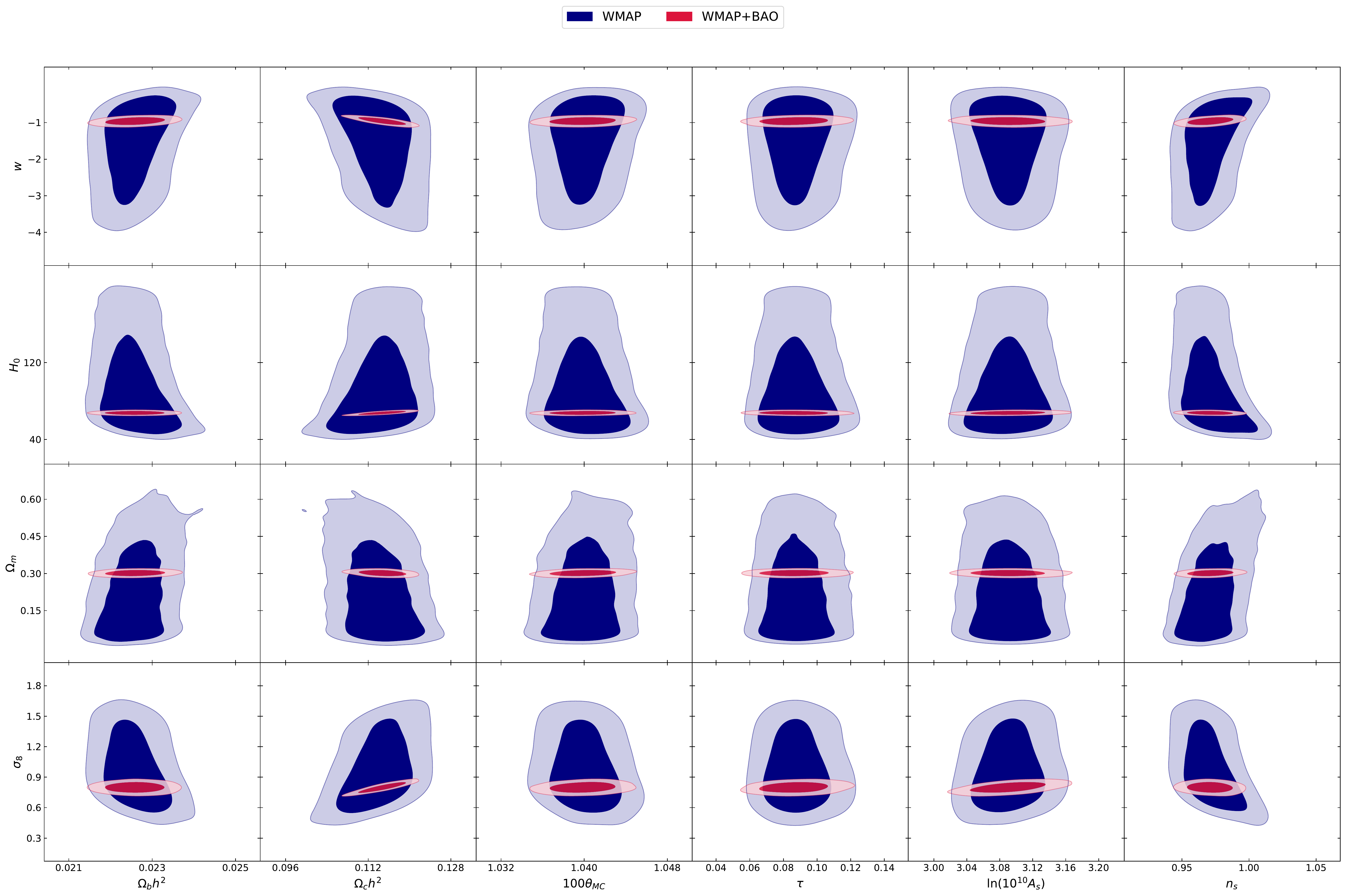}
\caption{Likelihood contours (68\% and 95\%) of cosmological parameters in a flat $w$CDM model derived from WMAP and WMAP + BAO respectively.}
\label{fig:wcdm}
\end{figure}

FIG. \ref{fig:lcdm} and FIG. \ref{fig:wcdm} shows 2-dimensional marginalized constraints on
the six MCMC sampling parameters of the $\Lambda$CDM model and $w$CDM model
used to explore the posterior of parameters,
and plotted against the following derived parameters
(the Hubble constant $H_0$, matter density parameter $\Omega_m$
and late-time clustering amplitude $\sigma_8$).
Our results are based on 9-year WMAP (TT,TE,EE+lensing).
Here we plot the results using the combined datasets of 6dF+MGS+DR12(9-zbin)+DR14 (labeld by BAO).
The blue contours show the constraints using 9-year WMAP data alone,
and the red contours include BAO data sets (WMAP+BAO).
It is easy to see the BAO method is sensitive to the change of $H_0$, $\Omega_m$ and $\sigma_8$,
so it can be effectively improve the constraint of WMAP-only.

\begin{figure}[!htbp]
\centering
\includegraphics[width=.35\textwidth]{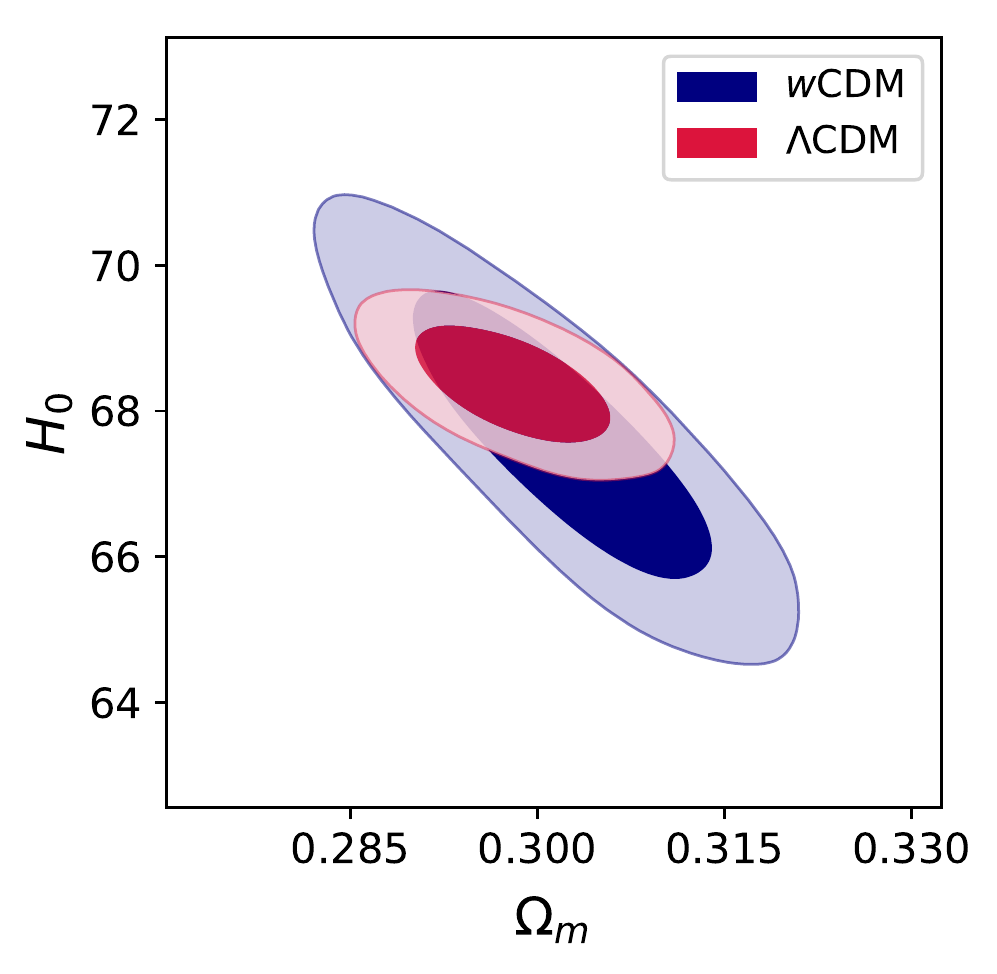}
\caption{Confidence contours for $\Omega_m$-$H_0$ in $\Lambda$CDM and $w$CDM model using WMAP+BAO data sets.}
\label{fig:omegam_h0}
\end{figure}

\begin{figure}[!htbp]
\centering
\includegraphics[width=.5\textwidth]{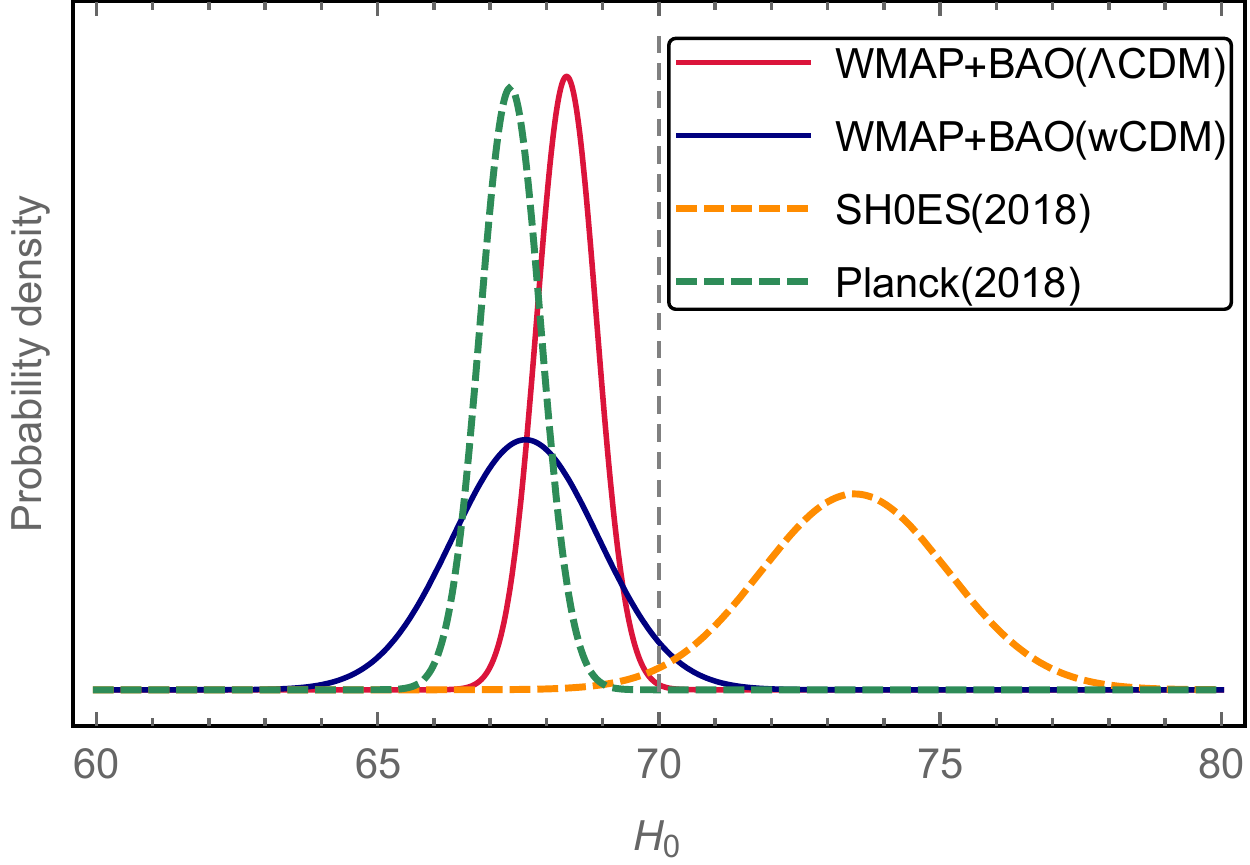}
\caption{Marginalized $H_0$ constraints for the $\Lambda$CDM and $w$CDM model
and comparison of the SH0ES and Planck measurements.}
\label{fig:h0}
\end{figure}

FIG. \ref{fig:omegam_h0} presents 68\% and 95\% likelihood contours
for the $\Omega_m$-$H_0$ plane for the WMAP+BAO data sets.
The red contours correspond to a flat $\Lambda$CDM model
and the blue contours correspond to the $w$CDM model.
FIG. \ref{fig:h0} shows the marginalized likelihood distribution of $H_0$ and summarized the $H_0$ measurements from other two methods.
Blue line and red lines show constraints in $\Lambda$CDM and $w$CDM model respectively
using 9-year WMAP data and BAO data.
Clearly, adding the recent BAO as a complementary to WMAP,
our measured $H_0$ results consistent with the recent data on CMB constraints from Planck (2018),
which prefer a value lower than 70 km s$^{-1}$ Mpc$^{-1}$.
These shows in 3.1 and $3.5\sigma$ tension with local measurements of Riess et al. (2018)
in $\Lambda$CDM and $w$CDM framework, respectively.

Table \ref{tab:result} gives marginalized parameter constraints from the WMAP CMB spectra
with and without BAO.
Parameter 68\% intervals in the $\Lambda$CDM model and $w$CDM model
from WMAP CMB power spectra in combination with BAO.
The first group is the base six parameters in $\Lambda$CDM model,
which are sampled in the MCMC analysis.
The second group lists the representative derived parameters ($H_0$, $\Omega_m$ and $\sigma_8$).
The third group shows the $\chi^2$ of WMAP and each BAO data sets.
The column labeled 'WMAP' is 9-year WMAP only.
The first two columns give results of six parameter $\Lambda$CDM
from 9-year WMAP data, with and without BAO measurements.
The last two columns give results in $w$CDM framework from WMAP data only and when BAO are added.
Adding BAO measurements to WMAP, we constrain cosmological parameters
$\Omega_m=0.298\pm0.005$, $H_0=68.36^{+0.53}_{-0.52}$ km s$^{-1}$ Mpc$^{-1}$ (0.78 \% precision),
$\sigma_8=0.8170^{+0.0159}_{-0.0175}$ for a flat $\Lambda$CDM model,
and $\Omega_m=0.302\pm0.008$, $H_0=67.63\pm1.30$ km s$^{-1}$ Mpc$^{-1}$ (1.93 \% precision),
$\sigma_8=0.7988^{+0.0345}_{-0.0338}$ for a flat wCDM model.
The combined constraint on $w$ from WMAP+BAO in a flat $w$CDM model is $w=-0.96\pm0.07$.
Compared with the WMAP alone analysis, the WMPA+BAO analysis reduces the error bar by
75.4\% in $\Lambda$CDM model and 95.3\% in $w$CDM model.

\begin{table}[!htbp]
\centering
\begin{tabular}{l|c|c|c|c}
  \hline
  \hline
  \multirow{2}*{Parameter} & \multicolumn{2}{c|}{$\Lambda$CDM} & \multicolumn{2}{c}{$w$CDM} \\
  \cline{2-5} & WMAP & WMAP+BAO & WMAP & WMAP+BAO \\
  \hline
  $\Omega_b h^2$             & $0.02264 \pm 0.00050$           & $0.02248 \pm 0.00043$
                             & $0.02261^{+0.00049}_{-0.00057}$ & $0.02260 \pm 0.00046$ \\
  $\Omega_c h^2$             & $0.1136^{+0.0048}_{-0.0046}$    & $0.1162 \pm 0.0018$
                             & $0.1140^{+0.0048}_{-0.0049}$    & $0.1148 \pm 0.0030$ \\
  $100 \theta_{\rm MC}$      & $1.04007 \pm 0.00221$           & $1.03961^{+0.00201}_{-0.00203}$
                             & $1.04002^{+0.00230}_{-0.00227}$ & $1.03998^{+0.00215}_{-0.00195}$ \\
  $\tau$                     & $0.0891^{+0.0123}_{-0.0148}$    & $0.0856^{+0.0112}_{-0.0132}$
                             & $0.0885^{+0.0121}_{-0.0148}$    & $0.0874^{+0.0120}_{-0.0138}$ \\
  $\ln(10^{10} A_s)$         & $3.092 \pm 0.029$               & $3.092^{+0.026}_{-0.030}$
                             & $3.091^{+0.029}_{-0.030}$       & $3.092^{+0.026}_{-0.030}$ \\
  $n_s$                      & $0.9728^{+0.0122}_{-0.0139}$    & $0.9678^{+0.0098}_{-0.0099}$
                             & $0.9722^{+0.0126}_{-0.0166}$    & $0.9713^{+0.0111}_{-0.0112}$ \\
  $w$                        & -                               & -
                             & $-1.59^{+1.26}_{-0.53}$         & $-0.96 \pm 0.07$ \\
  \hline
  $H_0$                      & $69.65^{+2.08}_{-2.37}$         & $68.36^{+0.53}_{-0.52}$
                             & $94.49^{+16.48}_{-46.45}$       & $67.63 \pm 1.30$ \\
  $\Omega_m$                 & $0.284^{+0.025}_{-0.028}$       & $0.298 \pm 0.005$
                             & $0.230^{+0.081}_{-0.197}$       & $0.302 \pm 0.008$ \\
  $\sigma_8$                 & $0.8075^{+0.0242}_{-0.0222}$    & $0.8170^{+0.0159}_{-0.0175}$
                             & $0.9697^{+0.2024}_{-0.3639}$    & $0.7988^{+0.0345}_{-0.0338}$ \\
  \hline
  $\chi^2_{\rm WMAP}$        & $7564.0676$                     & $7563.3854$
                             & $7564.8138$                     & $7563.4906$ \\
  $\chi^2_{\rm 6 DF}$        & -                               & $0.0383$
                             & -                               & $0.0628$ \\
  $\chi^2_{\rm MGS}$         & -                               & $2.1741$
                             & -                               & $1.8435$ \\
  $\chi^2_{\rm DR12(9zbin)}$ & -                               & $13.4406$
                             & -                               & $14.4571$ \\
  $\chi^2_{\rm DR14}$        & -                               & $0.0280$
                             & -                               & $0.0165$ \\
  $\chi^2_{\rm BAO}$         & -                               & $15.6810$
                             & -                               & $16.3799$ \\
  \hline
  \hline
\end{tabular}
\caption{Parameter constraints in $\Lambda$CDM and $w$CDM from the WMAP with and without BAO.}
\label{tab:result}
\end{table}

\section{Summary and discussion}
\label{sd}

In this paper, we determine the Hubble constant $H_0$
using the cosmic microwave background (CMB) data from WMAP
and the latest baryon acoustic oscillation (BAO) measurements
in a spatially flat $\Lambda$CDM and $w$CDM cosmology.
Adding BAO measurements to WMAP, we constrain cosmological parameters
$\Omega_m=0.298\pm0.005$, $H_0=68.36^{+0.53}_{-0.52}$ km s$^{-1}$ Mpc$^{-1}$ (0.78 \% precision),
$\sigma_8=0.8170^{+0.0159}_{-0.0175}$ in a flat $\Lambda$ cold dark matter ($\Lambda$CDM) model,
and $\Omega_m=0.302\pm0.008$, $H_0=67.63\pm1.30$ km s$^{-1}$ Mpc$^{-1}$ (1.93 \% precision),
$\sigma_8=0.7988^{+0.0345}_{-0.0338}$ in a flat $w$CDM model.
The combined constraint on $w$ from CMB and BAO for a flat $w$CDM model is $w=-0.96\pm0.07$.
By adding the recent BAO as a complementary to WMAP,
our measured $H_0$ results consistent with the recent data on CMB constraints from Planck (2018),
which prefer a value lower than 70 km s$^{-1}$ Mpc$^{-1}$.
These shows in 3.1 and $3.5\sigma$ tension with local measurements of Riess et al. (2018)
in $\Lambda$CDM and $w$CDM framework, respectively.
Compared with the WMAP alone analysis, the WMPA+BAO analysis reduces the error bar by
75.4\% in $\Lambda$CDM model and 95.3\% in $w$CDM model.

Our results indicate that the combination of WMAP and BAO datasets gives a tight constraint on the Hubble constant comparable to that adopting Planck data. In order to soften the model-dependent constraint using CMB and BAO data, we also extend our analysis to more general dark energy model ($w$CDM cosmology), but there is still a significant tension between the global fitting  CMB and BAO datasets and local determination.

\vspace{5mm}
\noindent {\bf Acknowledgments}

We acknowledge the use of HPC Cluster of ITP-CAS.
This work is supported by grants from NSFC
(grant No. 11575271, 11690021, 11747601),
the Strategic Priority Research Program of Chinese Academy of Sciences
(Grant No. XDB23000000), Top-Notch Young Talents Program of China,
and Key Research Program of Frontier Sciences of CAS.



\end{document}